\documentclass[12pt]{article}

\usepackage{times}
\usepackage{color}
\usepackage{url}
\newenvironment{sciabstract}{%
\begin{quote} \bf}
{\end{quote}}

\usepackage{indentfirst}
\usepackage{latexsym}
\usepackage{amsmath}
\usepackage{bm} % bold math
\usepackage{pifont}
\usepackage{latexsym}
\usepackage{amsmath, amsthm, amssymb, pifont, wasysym}
\usepackage{dcolumn} % Align table columns on decimal point
\usepackage{color}
\usepackage{graphicx}
\usepackage[labelfont=bf]{caption}
\usepackage[labelsep=period]{caption}

\newcommand {\TC}{$T_{\mathrm{C}}$}

\newcommand {\ECS}{EuCd$_2$Sb$_2$}

\newcommand {\SRO}{SrRuO$_3$}
\newcommand {\STO}{SrTiO$_3$}

\newcommand {\rhoxx}{${\rho_{\mathrm{xx}}}$}
\newcommand {\rhoyx}{${\rho_{\mathrm{yx}}}$}
\newcommand {\rhoyxzero}{${\rho_{\mathrm{yx, 0T}}}$}

\newcommand {\sxy}{${\sigma_{\mathrm{xy}}}$}

\newcounter{lastnote}

\title{Spontaneous in-plane anomalous Hall response\\ observed in a ferromagnetic oxide}

\author
{Shinichi Nishihaya,$^{1}$ Yuta Matsuki,$^{1}$ Haruto Kaminakamura,$^{1}$\\
 Yoshiya Murakami,$^{1}$ Hiroaki Ishizuka,$^{1}$ Masaki Uchida$^{1, \ast}$
\\ 
\\
\normalsize{$^{1}$Department of Physics, Institute of Science Tokyo, Tokyo 152-8551, Japan}\\
\normalsize{$^\ast$To whom correspondence should be addressed: E-mail: m.uchida@phys.sci.isct.ac.jp}
}
\date{}
\begin{document}
\baselineskip24pt
\maketitle 
\begin{sciabstract}
{\bf
Recent observation of anomalous Hall effect (AHE) induced by magnetic field or spin magnetization lying in the Hall deflection plane has sparked interest in diverse mechanisms for inducing the Hall vector component perpendicular to the applied magnetic field. Such off-diagonal coupling, which is strictly constrained by symmetry of the system, provides new degrees of freedom for engineering Hall responses. However, spontaneous response as extensively studied for out-of-plane AHE remains unexplored. Here we elucidate in-plane AHE in a typical ferromagnetic oxide {\SRO}. The (111)-orientated ultrathin films with in-plane easy axes of spin magnetization exhibit spontaneous AHE at zero field, which is intrinsically coupled to the in-plane spin magnetization and controllable via its direction. Systematic measurements by varying azimuthal and polar field angles further reveal complex Hall responses shaped by higher-order terms allowed by trigonal distortion of the films. Our findings highlight versatile and controllable in-plane Hall responses with out-of-plane orbital ferromagnetism.
}
\end{sciabstract}

\clearpage
\section*{Introduction}
Manipulation of spin and orbital magnetization by magnetic field is fundamental to diverse magnetotransport phenomena in magnets. The primary effect of the magnetic field is the Zeeman-type coupling, where magnetic moments couple diagonally to the applied field \cite{Zeeman}. Recently, on the other hand, observation of Hall responses under the field applied within the Hall deflection plane has been reported (Fig. 1A) \cite{iAHEexp1,iAHEexp2,iAHEexp3_111,iAHEtheexp_Fe3Sn2,iAHEtheexp_ECS,iAHEtheexp_Fe}. In contrast to the conventional Hall effect induced by the out-of-plane magnetic field \cite{Hall1,AHEreview}, in-plane field induced anomalous Hall effect (in-plane AHE) can be interpreted as generation of the Hall vector component perpendicular to the in-plane magnetic field. Namely, it is an off-diagonal response to the magnetic field. 

Following the theoretical proposals of in-plane AHE and its quantization in topological materials \cite{iAHEthe1_Q,iAHEthe2_Q,iAHEthe3_Q,iAHEthe4_Q,iAHEthe7_Q_111,iAHEthe5,iAHEthe6,iAHEthe9}, experimental observations of in-plane AHE in various materials have invoked interest in detailed mechanisms for realizing such off-diagonal responses  \cite{iAHEexp1,iAHEexp2,iAHEexp3_111,iAHEtheexp_Fe3Sn2,iAHEtheexp_ECS,iAHEtheexp_Fe}. Among them, the in-plane AHE observed for Weyl semimetals Fe${_3}$Sn${_2}$ \cite{iAHEtheexp_Fe3Sn2} and {\ECS} \cite{iAHEtheexp_ECS}, which exhibits three-fold symmetry for the in-plane field rotation, has been understood as a consequence of band structure and quantum geometry modulations induced by the in-plane magnetic field. The in-plane field induced modulations are of purely intrinsic origins and thus strictly follows the crystal symmetry  \cite{iAHEsym,iAHEsym2}, exhibiting the unique field direction dependence (Fig. 1B). In particular, out-of-plane Weyl points shifting by the in-plane field corresponds to manifestation of orbital magnetization in the out-of-plane direction \cite{orbitalmagterm1,orbitalmag_rev,orbitalmagterm2}. 

So far research on in-plane AHE has been limited to non-magnetic \cite{iAHEexp1,iAHEexp2,iAHEexp3_111}, antiferromagnetic \cite{iAHEtheexp_ECS} or soft-magnetic materials \cite{iAHEtheexp_Fe3Sn2,iAHEtheexp_Fe} with negligible magnetic hysteresis. As shown in Fig. 1C, hard ferromagnets potentially host anomalous Hall conduction and out-of-plane orbital ferromagnetism at zero field. Such a spontaneous response involving the off-diagonal coupling between orbital and spin degrees of freedom provides new opportunities for engineering the Hall transport. On the other hand, spin canting or reorientation derived by magnetocrystalline or shape magnetic anisotropy in ferromagnets may complicate observations by inducing the out-of-plane spin magnetization component and conventional out-of-plane AHE. From this viewpoint, a ferromagnet with in-plane easy axes is highly desired for elucidating the pure response coupled to the in-plane spin magnetization.

Here, we study spontaneous in-plane AHE observed in films of ferromagnetic perovskite oxide {\SRO}. The band structure of {\SRO} hosting several Weyl point pairs near the Fermi level \cite{AHEexp1_SROWeyl1,SROWeyl2} is expected to be beneficial for realizing large in-plane AHE. (111)-oriented ultrathin {\SRO} films are epitaxially grown on the (111) {\STO} substrate with threefold rotational symmetry about the out-of-plane direction (see Supplementary Note S1 for structural characterization). A giant in-plane AHE comparable to the out-of-plane AHE is observed in the {\SRO} ultrathin films, where the anomalous Hall conduction persists at zero field and can be controlled via the in-plane spin magnetization direction. Detailed measurements by varying azimuthal and polar angles of the field confirm that the {\SRO} ultrathin films possess in-plane easy axes, and reveal that the observed in-plane AHE is an off-diagonal response intrinsically coupled to the in-plane spin magnetization through higher-order terms allowed by trigonal distortion of the films.

\section*{Results}

\noindent {\bf Out-of-plane and in-plane AHE}

Figure 2A shows conventional Hall resistivity {\rhoyx} measured for a $(111)$ {\SRO} film with thickness of 4.1 nm (sample A) with sweeping the out-of-plane magnetic field at 2 K. The Curie temperature {\TC} of this sample is determined to be 130 K, which is slightly lower than the bulk value but consistent with previous studies on $(111)$ {\SRO} thin films (see Supplementary Note S2 for other fundamental transport) \cite{SRO111_1, SRO111_2, SRO111_3}. {\rhoyx} shows a large hysteresis loop characteristic to hard ferromagnets, also in good agreement with the previous studies  \cite{SRO111_1, SRO111_2, SRO111_3}. The slope above the coercive field is roughly 0.03 $\mu\Omega\mathrm{cm}/\mathrm{T}$.

AHE has not been fully examined in ferromagnets with in-plane spin magnetization. As shown in Fig. 2B, however, {\rhoyx} taken with sweeping the in-plane [$11\bar{2}$] field at $\varphi=0^{\circ}$ exhibits significantly large values comparable to the out-of-plane scan, and continues to increase above the coercive field with a similar slope of 0.03 $\mu\Omega\mathrm{cm}/\mathrm{T}$. Moreover, {\rhoyx} in the in-plane scan also remains finite at zero magnetic field accompanied with large hysteresis. Importantly, the present {\SRO} ultrathin films possess in-plane easy axes as shown later with the field angle dependence, and the observed AHE cannot be explained by out-of-plane canting of spin magnetization. The observation rather highlights the importance of out-of-plane orbital magnetization coupled to the in-plane spin magnetization.

{\rhoyx} taken at $\varphi = 60^{\circ}$ shows similar in-plane field dependence with signs opposite to the $\varphi = 0^{\circ}$ case. These observations are also consistent with symmetry of the trigonally distorted $(111)$ {\SRO} thin films, which have a $C_3$ axis along the $[111]$ direction, $C_2$ axes along the $[1\bar{1}0]$ and its equivalent directions, and mirror planes on the $(1\bar{1}0)$ and its equivalent planes. This allows the emergence of finite in-plane AHE with opposite signs centered at $\varphi = 0^{\circ}, 120^{\circ}, 240^{\circ}$ and $\varphi = 60^{\circ}, 180^{\circ}, 300^{\circ}$  \cite{iAHEsym,iAHEsym2}. 

\noindent {\bf Azimuthal angle dependence}

Figure 3A demonstrates $\varphi$ dependence of {\rhoyx}, taken with rotating the in-plane magnetic field on the $(111)$ plane for sample A. $\rho_{\mathrm{yx}}( \varphi )$ is derived by antisymmetrization of the raw data $\rho_{\mathrm{yx,raw}}$, as expressed by $\rho_{\mathrm{yx}}( \varphi )=(\rho_{\mathrm{yx,raw}}( \varphi )-\rho_{\mathrm{yx,raw}}( \varphi +180^{\circ}))/2$. When the field is above the in-plane coercive field of about 3 T, {\rhoyx}($\varphi$) exhibits a sinusoidal curve with three-fold symmetry, which is similar to the in-plane AHE reported for antiferromagnets and soft ferromagnets  \cite{iAHEtheexp_Fe3Sn2, iAHEtheexp_ECS, iAHEtheexp_Fe}. Its magnitudes and signs at $\varphi = 0^{\circ}$ and $60^{\circ}$ are also consistent with the in-plane field scan data.

To investigate $\varphi$ dependence of the Hall resistivity at zero field {\rhoyxzero}, we repeatedly performed a procedure at each $\varphi$, which involves increasing the in-plane field to 9 T, returning it to 0 T, and then measuring the Hall resistivity. Here $\rho_{\mathrm{yx,0T}}( \varphi )$ is obtained by antisymmetrization of a pair of raw data $\rho_{\mathrm{yx,0T,raw}}( \varphi )$ and $\rho_{\mathrm{yx,0T,raw}}( \varphi +180^{\circ})$. As shown in Fig. 3B, {\rhoyxzero} exhibits a rather square-wave curve with three-fold symmetry for rotation of the polarizing field, taking values consistent with the field scans at $\varphi = 0^{\circ}$ and $60^{\circ}$ in Fig. 2B. {\rhoyxzero} decreases with increase in temperature and disappears at {\TC}, confirming that this response is indeed related to the ferromagnetic ordering in {\SRO} (see Supplementary Note S3 for its detailed temperature dependence). Similar square-wave curves are reproducibly confirmed in {\rhoyxzero} taken for another (111) {\SRO} film with thickness of 7.0 nm (sample B) shown in Fig. 3C, regardless of the current direction on the plane. Difference of the $\varphi$ dependence between the sinusoidal wave at 9 T and the square-like wave at 0 T suggests the presence of magnetic anisotropy with in-plane easy axes pointing to the $[11\bar{2}]$ and its equivalent directions.

\noindent {\bf Polar angle dependence}

To further clarify the magnetic anisotropy near zero field, we present $\theta$ dependence of {\rhoyxzero} measured on the $(\bar{1}10)$ or $\varphi=0^{\circ}$ plane in Fig. 4A. Here $\rho_{\mathrm{yx,0T}}( \theta )$ is derived by antisymmetrization of $\rho_{\mathrm{yx,0T,raw}}( \theta )$ and $\rho_{\mathrm{yx,0T,raw}} ( \theta+180^{\circ})$. {\rhoyxzero}($\theta$) exhibits a plateau structure not only around the out-of-plane $[111]$ and $[\bar{1}\bar{1}\bar{1}]$ directions but also over the range including the in-plane $[11\bar{2}]$ direction. In addition,  the $\theta$ scans taken at various magnetic fields (see Figure Supplementary Figure S4) reveal that {\rhoyx}($\theta$) exhibits a pronounced hysteresis loop around the $[00\bar{1}]$ and $[11\bar{1}]$ directions while the hysteresis loop is closed around the in-plane $[11\bar{2}]$ direction. All of these observations evidence that the {\SRO} ultrathin film possesses relatively strong in-plane shape magnetic anisotropy in addition to $\langle111\rangle$ magnetocrystalline anisotropy, realizing a ferromagnetic state with the spins aligned to the in-plane $[11\bar{2}]$ direction when the $\theta$ is close to $90^\circ$. Therefore, the observed spontaneous in-plane AHE is not due to out-of-plane canting of the spin magnetization, and it can be regarded as out-of-plane orbital ferromagnetism off-diagonally coupled to the in-plane spin magnetization. 

Figure 4B presents $\theta$ scans measured at 9 T for the sweep from $[\bar{1}\bar{1}\bar{1}]$ toward $[111]$ and its reverse on the $\varphi=0^{\circ}$ plane (See Supplementary Figure S5 for $\theta$ scans performed on the $\varphi=60^{\circ}$ plane). Figure 4C compares $\theta$ scans measured at various magnetic fields for a forward sweep from $[\bar{1}\bar{1}\bar{1}]$ to $[111]$ (see also Supplementary Figures S4 for both the sweeps at various fields). While the effect of in-plane magnetic anisotropy is pronounced at the low fields, {\rhoyx}($\theta$) measured at high fields such as above 6 T exhibits negligibly small hysteresis, indicating that the spin magnetization almost follows the applied field direction during the sweep. {\rhoyx}($\theta$) at high fields exhibits nonmonotonic dependence accompanied by local minimum and maximum around $[11\bar{1}]$ and $[00\bar{1}]$ directions, respectively. In particular, the local maximum around $[11\bar{1}]$ exhibiting a positive sign, which is opposite to the sign at $[111]$. This sign change feature strongly indicates the necessity of considering a higher-order effect due to magnetic field $B$  \cite{iAHEtheexp_ECS} or magnetization $M$  \cite{iAHEtheexp_Fe}, which can be effectively regarded as generation of the out-of-plane orbital magnetization component by the in-plane field.

\section*{Discussion}

As the Hall effect is a field-odd response, $B_{i}B_{j}B_{k}$ or $M_{i}M_{j}M_{k}$ with $i, j, k$ Cartesian indices can be the leading term of the higher-order effect. By introducing the direction cosines $\alpha_{i}$ and $\beta_{i}$ and unit vectors $\bm{e}_{i}$ along the principal axes, a general expansion of the Hall conductivity on the $xy$ plane perpendicular to $\hat{\bm{z}}=\beta_{i} \bm{e}_{i}$ in cubic systems is expressed as
\begin{equation}
\sigma_{xy, \rm{cubic}}(\bm{B})=\sigma^{(1)}B\alpha_{i}\beta_{i}+\sigma^{(3)}B^3\alpha_{i}^3\beta_{i}+\sigma^{(5)}B^5\alpha_{i}^5\beta_{i}+\cdots
\end{equation}
or
\begin{equation}
\sigma_{xy, \rm{cubic}}(\bm{M})=\sigma^{(1)}M\alpha_{i}\beta_{i}+\sigma^{(3)}M^3\alpha_{i}^3\beta_{i}+\sigma^{(5)}M^5\alpha_{i}^5\beta_{i}+\cdots
\end{equation}
for magnetic field vector $\bm{B}=B\alpha_{i} \bm{e}_{i}$ or magnetization vector $\bm{M}=M\alpha_{i} \bm{e}_{i}$. By introducing anisotropy terms, the anomalous Hall conductivity outside the hysteresis loop can be expanded with respect to the magnetic field in the end. According to Eq. (1), {\rhoyx} on the $(111)$ plane is usually dominated by the field-linear term proportional to $B_{[100]} + B_{[010]} +B_{[001]}$, or equivalently, to $B_{[111]}$; namely, {\rhoyx} changes proportionally with $B \cos\theta$, resulting in the absence of in-plane AHE. In cases with significant contributions from higher-order terms, on the other hand, {\rhoyx} manifests a characteristic $\theta$ dependence reflecting the crystal symmetry, while it appears always with three-fold rotational symmetry in the $\varphi$ scan.

Figure 4D shows simulated $\theta$ dependence of the anomalous Hall conductivity {\sxy} on the $\varphi = 0^{\circ}$ plane considering different higher-order terms. The third-order term in {\rhoyx} on the cubic $(111)$ plane is proportional to $B_{[100]}^3 + B_{[010]}^3 +B_{[001]}^3$, leading to a nonmonotonic $\theta$ dependence with a finite value at $\theta = 90^{\circ}$ (black curve in Fig. 4D). On the other hand, the experimentally observed behavior shown in Fig. 4B cannot be fitted well with the $B_{[100]}^3 + B_{[010]}^3 +B_{[001]}^3$ term. This suggests that it may be necessary to consider the effect of trigonal distortion which breaks the three-fold rotational symmetry about the $[\bar{1}11]$, $[1\bar{1}1]$, and $[11\bar{1}]$ directions (see Supplementary Note S1 for structural characterization). This allows other third-order terms such as $B_{[100]} B_{[010]} B_{[001]}$, which actually involves the local maximum of {\sxy} at $[11\bar{1}]$ with a sign opposite to $[111]$ on the $\varphi = 0^{\circ}$ plane (red curve in Fig. 4D). From the same discussion, the spontaneous Hall response at zero field can be understood based on the higher order terms of the spin magnetization $M$.

In summary, we demonstrate spontaneous in-plane AHE in (111)-oriented ultrathin films of a hard ferromagnet {\SRO}. Reflecting the crystal symmetry, the in-plane AHE emerges with three-fold symmetry for the field rotation on the Hall deflection plane. Off-diagonally coupled to the in-plane spin magnetization, the spontaneous anomalous Hall conduction and associated out-of-plane orbital magnetization persist even at zero field, and their sign can be switched via coupling to the in-plane spin magnetization. Moreover, the polar angle scans reveal the peculiar nonmonotonic behavior which suggests the contribution of the higher-order terms allowed under the trigonal distortion. While it is difficult to quantify the contribution from each term, our observations highlight the unique and spontaneous appearance of in-plane AHE reflecting the crystal symmetry, magnetic hysteresis, and magnetic anisotropy in the conventional ferromagnet. The present work broadens the horizons of Hall physics by demonstrating essential control of the Hall conduction by the in-plane magnetic field and its history.

\section*{Materials and Methods}

\noindent {\bf Epitaxial film growth}

(111)-oriented {\SRO} films were grown on (111) {\STO} substrates in an Eiko EB-9000S oxide molecular beam epitaxy chamber equipped with a semiconductor-laser heating system  \cite{SRO113MBE, SRO327MBE}. {\STO} substrates were annealed at 870 $^\circ$C prior to the growth, and then {\SRO} films were grown at 650 $^\circ$C by supplying 4N Sr from a conventional Knudsen cell, 3N5 Ru from an electron beam evaporator, and O$_3$ (60\%) + O$_2$ (40\%) from a Meidensya MPOG-RDE01C ozone generator. The film thickness was typically designed at about 4 nm.

\noindent {\bf Magnetotransport measurements}

Longitudinal resistivity {\rhoxx} and Hall resistivity {\rhoyx} on Hall bar devices were measured using the conventional low-frequency lock-in technique. Field angle dependences of {\rhoxx} and {\rhoyx} up to 9 T were measured using a sample rotator in a Cryomagnetics cryostat system equipped with a superconducting magnet. The magnetotransport measurements were performed with changing the magnetic field direction within the (111) plane with an azimuthal angle $\varphi$ (measured from [11$\bar{2}$]), and also from the out-of-plane [111] direction toward an in-plane one with a polar angle $\theta$ (measured from [111]).

%\bibliography{scibib}

%\section*{References and Notes}

%%%%%%%%%%%%%%%%%%%%%%%%%

\section*{Acknowledgments}
\textbf{Funding:} This work was supported by JST FOREST Program Grant Number JPMJFR202N and PRESTO Program Grant Number JPMJPR2452, by JSPS KAKENHI Grant Numbers JP22K18967, JP22K20353, JP23K13666, JP23K03275, JP24H01614, and JP24H01654 from MEXT, Japan, by Murata Science and Education Foundation, Japan, and by STAR Award funded by the Tokyo Tech Fund, Japan.
\textbf{Author contributions:} M.U. conceived the project and designed the experiments. Y.Ma. and H.K. grew films and performed transport measurements with S.N. and Y.Mu. Y.Ma., H.K., and S.N. analyzed the data and M.U. and S.N. wrote the manuscript with input from all authors. H.I. jointly discussed the results. All authors have approved the final version of the manuscript.
\textbf{Competing interests:} The authors declare that they have no competing interests.
\textbf{Data and materials availability:} All data needed to evaluate the conclusions in the paper are present in the paper and/or the Supplementary Materials.

\clearpage

%%%%%%%%%%%%%%%%%%%%%%%%%
%Figures
%%%%%%%%%%%%%%%%%%%%%%%%%

\begin{figure}[t]
\centering
\includegraphics[width=13.5cm]{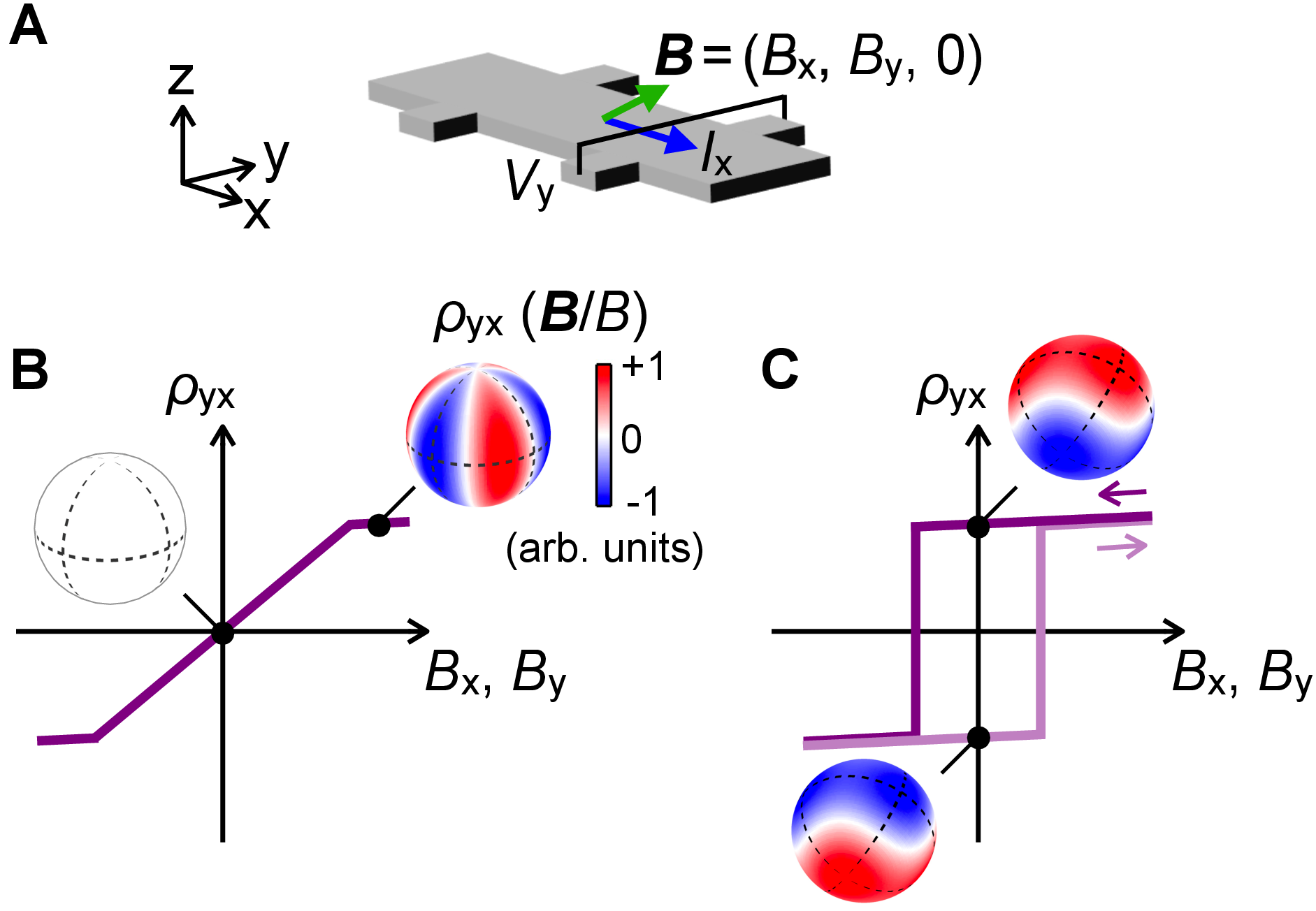}
\caption{\label{fig1}\textbf{Anomalous Hall effect by in-plane magnetic field.}
(A) Measurement configuration of anomalous Hall effect (AHE) under the in-plane magnetic field. A voltage $V_{\mathrm{y}}$ transverse to an electric current $I_{\mathrm{x}}$ can be generated depending not only on the value but also on the history of the in-plane magnetic fields $B_{\mathrm{x}}$ and $B_{\mathrm{y}}$. (B) For antiferromagnets, the Hall resistivity {\rhoyx} becomes zero at zero magnetic field, even in the case that large {\rhoyx} with three-fold rotational symmetry around the $\mathrm{z}$ direction is induced by the in-plane field. A case of (001)-oriented trigonal systems is shown as an example of the field direction dependence of {\rhoyx}. (C) In hard ferromagnets, it is expected that {\rhoyx} spontaneously emerges even after turning off the in-plane magnetic field. A case of (111)-oriented cubic systems is exemplified.
}
\end{figure}

\begin{figure}[t]
\centering
\includegraphics[width=15.5cm]{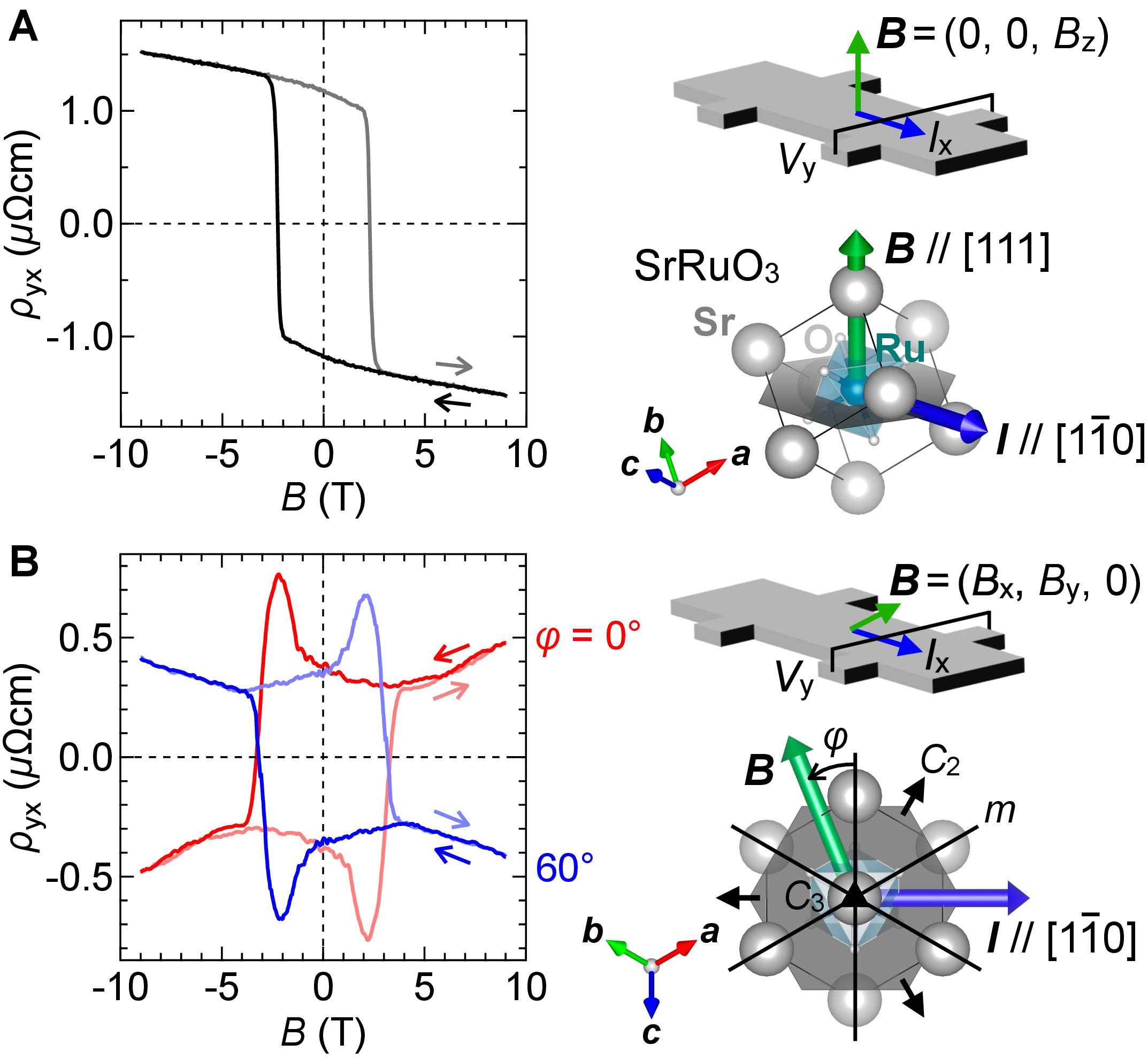}
\caption{\label{fig:fig2}\textbf{Out-of-plane and in-plane AHE with a hysteresis loop.}
(A) Hall resistivity {\rhoyx} of a (111) {\SRO} film with thickness of 4.1 nm (sample A), taken with sweeping the out-of-plane magnetic field along the $[111]$ direction at 2 K. Schematic illustration of the measurement configuration and its correspondence to the {\SRO} crystal structure are shown in the right panel. (B) {\rhoyx} measured with sweeping the in-plane field at azimuthal angles $\varphi = 0^{\circ}$ and $60^{\circ}$, where $\varphi$ is measured from the $[11\bar{2}]$ direction. Illustration of the in-plane measurement configuration and its correspondence to the crystal structure are similarly shown, together with fundamental symmetry elements of the $C_2$ and $C_3$ rotation axes and the mirror planes. 
}
\end{figure}

\begin{figure}[t]
\centering
\includegraphics[width=11.5cm]{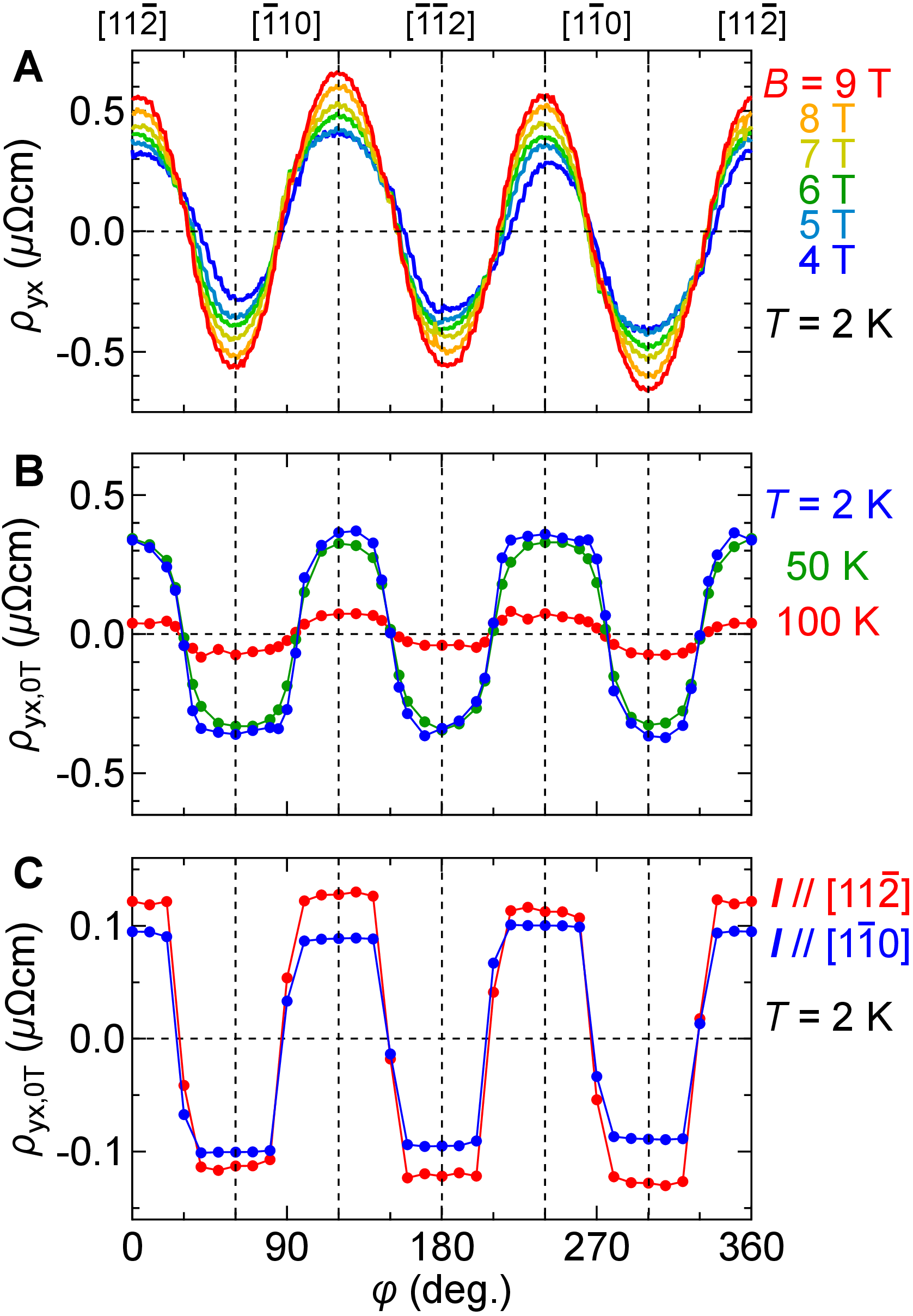}
\caption{\label{fig:fig3}\textbf{In-plane AHE at zero magnetic field.}
(A) {\rhoyx} of the {\SRO} film (sample A), taken with continuously changing $\varphi$ of various in-plane magnetic fields at 2 K. (B) Hall resistivity at zero magnetic field {\rhoyxzero}, measured after applying the in-plane field of 9 T and then lowering it to 0 T at each $\varphi$. The measurements were performed at 2, 50, and 100 K. (C) {\rhoyxzero} measured for another (111) {\SRO} film with thickness of 7.0 nm (sample B) after the same magnetization process at 2 K. The measurements were performed on two Hall bar devices where the electric current flows along the $[1\bar{1}0]$ and $[11\bar{2}]$ directions, respectively.
} 
\end{figure}

\begin{figure}[t]
\centering
\includegraphics[width=15.0cm]{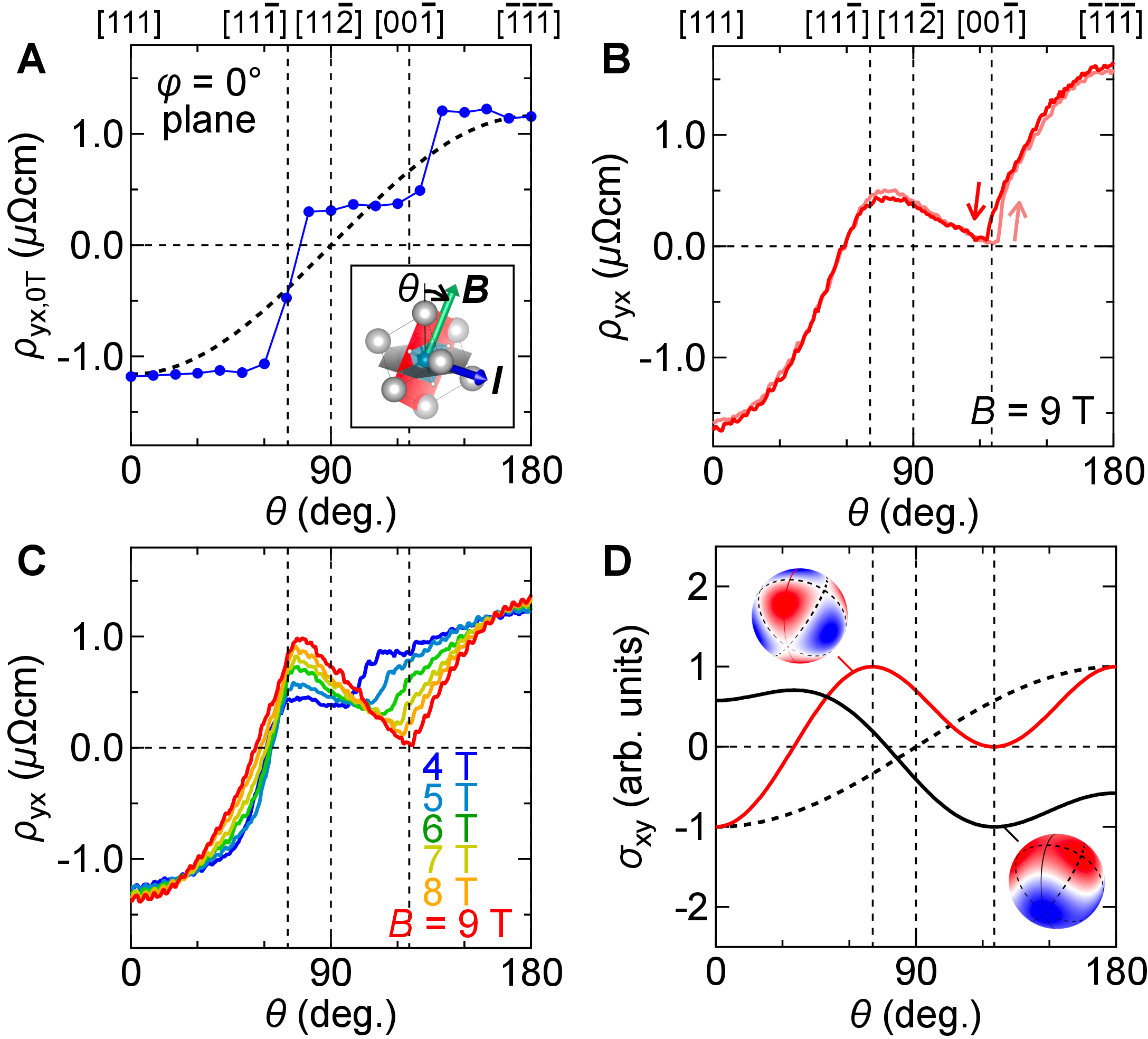}
\caption{\label{fig:fig4}\textbf{Nonmonotonic polar angle dependence of AHE.}
(A) {\rhoyxzero} of the {\SRO} film (sample A), measured after applying the field of 9 T and then lowering it to 0 T at each polar angle $\theta$ on the $\varphi = 0^{\circ}$ plane at 2 K. $\theta$ is measured from the out-of-plane $[111]$ direction. (B) $\theta$ dependence of {\rhoyx} measured at 9 T for forward and reverse sweeps between $[111]$ and $[\bar{1}\bar{1}\bar{1}]$. (C) Comparison of {\rhoyx}($\theta$) measured at various magnetic fields. Only the forward sweeps from $[\bar{1}\bar{1}\bar{1}]$ to $[111]$ are shown. (D) Simulated $\theta$ dependence of the anomalous Hall conductivity {\sxy} on the $\varphi = 0^{\circ}$ plane. Black and red solid curves represent {\sxy} proportional to the $B_{[100]}^3 + B_{[010]}^3 +B_{[001]}^3$ and $B_{[100]}B_{[010]}B_{[001]}$ terms, respectively, while the conventional $B_{[100]} + B_{[010]} +B_{[001]}$ term is shown by a dashed curve.
} 
\end{figure}

\clearpage

\end{document}